\documentclass[article,preprint,12pt,unsrt]{elsarticle}
\usepackage{a4wide}
\usepackage{natbib} 			
\usepackage{graphicx}
\usepackage{amsfonts}
\usepackage{amsmath}
\usepackage{booktabs}
\usepackage{epsfig}
\usepackage{stmaryrd}
\usepackage{hyperref}
\hypersetup{citecolor=black,linkcolor= black}

\begin{document}
\title{%Is $H_{xy}>(H_x+H_y)/2$ possible in the power-law cross-correlations setting?
Worldwide clustering of the corruption perception}
\author{Michal Paulus}
\author{Ladislav Kristoufek}
\ead{kristoufek@icloud.com}
\address{Institute of Economic Studies, Faculty of Social Sciences, Charles University, Opletalova 26, 110 00, Prague 1, Czech Republic
}

\begin{abstract}
We inspect a possible clustering structure of the corruption perception among 134 countries. Using the average linkage clustering, we uncover a well-defined hierarchy in the relationships among countries. Four main clusters are identified and they suggest that countries worldwide can be quite well separated according to their perception of corruption. Moreover, we find a strong connection between corruption levels and a stage of development inside the clusters. The ranking of countries according to their corruption perfectly copies the ranking according to the economic performance measured by the gross domestic product per capita of the member states. To the best of our knowledge, this study is the first one to present an application of hierarchical and clustering methods to the specific case of corruption.
\end{abstract}


\journal{Physica A}

\maketitle


\section{Introduction}

Corruption has been understood as an important aspect forming societies as well as economic connections for a long time. Recently, it has also become more evident that grasping corruption quantitatively can yield important additions to models in social sciences and especially economics. Various studies examine corruption in fields of microeconomics, macroeconomics, institutional economics or international trade and its influence on economic performance, efficiency and decision making. Corruption is thus in the centre of interest of many empirical and theoretical works \cite{Wei2000,Svensson2005,Shleifer1993}.

The most prominent topic regarding corruption is its impact on economic growth showing a negative influence of corruption on the growth \cite{Mauro1995,Mauro1997}. Another attractive domain of corruption effects is in the inequality studies, which show that corruption increases inequality \cite{Gupta2002}. The increased corruption also tends to boost public investments but these yield lower returns \cite{Haque2008}. The higher corruption not only lowers the public investment returns but also the level of private investments with a higher public-to-private investment ratio as a result \cite{Mauro1995,Davoodi1997,Tarhan2008}.

In general, the empirical studies reveal negative effects of corruption on economic performance in terms of lower growth, private investments, higher inequality or wasting of talents. Except for the studies concerning the macroeconomic performance, corruption studies are becoming more popular in the international trade area as well. The theoretical effect of corruption on mutual trade is not straightforward as it can have both negative and positive effects. In the case of an adverse impact, corruption can be viewed as an additional tax because of bribes, a complicated enforcement of mutual agreements or an institutional safety \cite{Horsewood2012}. The other perspective is completely opposite. Corruption helps to facilitate mutual business and to guarantee basic rights when the institutions are of a low quality.

When we abandon theoretical considerations and explore empirical literature, we cannot unanimously conclude that corruption has only a negative role in FDI (foreign direct investments) or trade, although it seems to be mostly the case \cite{Egger2006,Busse2007,Habib2001}. Studies disrupting the prevailing negative effects of corruption are present as well. Egger \& Winner \cite{Egger2005} show that corruption can stimulate FDI in the short as well as the long term. They argue that a corrupt country attracts investments from other corrupt countries. In other words, a country attracts investments from countries with a similar institutional quality \cite{Cuervo-Cazura2006}. Brada \textit{et al.} \cite{Brada2012} go further and they develop a theoretical model showing that the most beneficial position in the FDI engagement is to be a ``medium corrupt'' country. Such a country has advanced technologies, good institutions but also knowledge of the corruption culture. Hence they can be engaged in an FDI mediation with corrupt as well as non-corrupt countries. If a country is highly corrupt, it has a problem attracting investments from low corrupt countries and vice versa.

Such an ambiguous role of corruption is found in the international trade analyses also. Horsewood \& Voicu \cite{Horsewood2012} report adverse effects of corruption on mutual trade. On the other hand, de Jong \& Bogmans \cite{DeJong2011} find that corruption can stimulate imports under special circumstances. Some further results can be found in Refs. \cite{Dutt2010,Goel2011,Thede2012}.

Analysis of corruption has been of interest in the interdisciplinary research as well. Shao \textit{et al.} \cite{Shao2007} demonstrate a power-law functional dependence between corruption level and economic factors, such as country wealth and foreign direct investments per capita. Podobnik \textit{et al.} \cite{Podobnik2008} report that an increase of the corruption perception index leads to an increase of the annual GDP per capita growth rate between 1999 and 2004. In addition, the authors introduce a new measure to quantify the relative corruption between countries based on their respective wealth as measure by GDP per capita. Podobnik \textit{et al.} \cite{Podobnik2013} further propose a production function (divided into the private and public sectors) where GDP depends on market capitalization, the public or private section workforce and competitiveness level to quantify the public sector efficiency. It is demonstrated that the less corrupt countries receive more investments that the more corrupt ones. Finally Podobnik \textit{et al.} \cite{Podobnik2012} argue that it is improbable that a country would rapidly improve its corruption rank.

Even though the effects of corruption on various economic indicators and measures have been often investigated, an important message remains hidden beneath the results. The reported findings suggest that there are at least several groups of countries which are differentiated according to their corruption level and these groups tend to trade with each other more and they invest into similar countries as well. However, there is no study examining the actual clustering of countries according to their corruption or their corruption perception. Here, we aim to fill this gap in the literature by bridging an economic topic with tools of complex systems and interdisciplinary physics -- clustering analysis. In the following section, we describe the utilized methodology in some detail. The next section describes the analyzed dataset followed by results presentation and interpretation. The last section concludes. We find that countries indeed cluster according to their corruption levels and we identify four well-divided clusters. Moreover, these clusters nicely correspond to a stage of development of the member countries. The level of corruption is thus found to be tightly connected to country's development.
 
\section{Methodology}

The clustering analysis has now quite a long history in economic and financial applications mainly due a pioneering work of Mantegna \cite{Mantegna1999} utilizing minimum spanning trees and hierarchical trees to uncover a hierarchical structure in the US stocks. The methodology has then been applied to various markets such as stocks and stock indices \cite{Onnela2002,Onnela2003,Onnela2004,Bonanno2004,Tumminello2007,Brida2008,Eom2009,DiMatteo2010,Eom2010,Nobi2014}, foreign exchange rates \cite{Mizuno2006,Naylor2007,Jang2011,Keskin2011}, import/export networks \cite{Kantar2011}, interest rates \cite{Tabak2009}, and commodities \cite{Tabak2010,Kristoufek2012,Kristoufek2013}. Most of the studies focus on correlation analysis, using the correlation matrix as a starting point. The dissimilarity (distance) matrix is constructed using the correlations which then allows for the spanning and hierarchical trees construction. However, our analysis cannot use such an approach and we must apply an alternative one due to a specific nature of the corruption data and a different aim of the analysis.

We analyze a corruption clustering of various countries. For each country, we get a time series of the Freedom from Corruption (COR) index of the Heritage Foundation which ranges between 0 and 100 with higher values meaning lower corruption (more details are given in the next section). We are primarily interested in a common level of the index as well as in a co-movement between indices of different countries. Therefore, we cannot use a standard correlation procedure which demeans the series and thus the information about the corruption level is lost. Instead of a correlation coefficient, we utilize the simple Euclidean distance $d_{xy}$ between time series (or in general vectors) $\{x_t\}$ and $\{y_t\}$ defined as
\begin{equation}
d_{xy}=\sqrt{\sum_{t=1}^{T}{(x_t-y_t)^2}}
\end{equation}
where $T$ is the time series length. Even though there are various possibilities for defining distances \cite{Mardia1979,Borg1997}, we opt for the most basic Euclidean one\footnote{We have applied other distance metrics as well and the results remain qualitatively practically intact.}. We thus have a distance between time series as a measure of dissimilarity between corruption levels in pairs of countries forming a distance matrix $D$, which is symmetric with zeros on the diagonal.

For the hierarchical structure construction, we utilize the average linkage clustering approach (also known as UPGMA). The algorithm starts with nodes in a network treated as a separate cluster each. The closest pair is identified based on the lowest distance in the distance matrix $D$ (except for the diagonal elements). This closest pair forms a new cluster and it is substituted into the distance matrix instead of the original two elements. The new cluster is assigned the average of the distances of all elements of the newly formed cluster. This is repeated until we have a single cluster \cite{Sorensen1948,Everitt2001,Defays1977}. Such construction can be seen as a cautious one as it uses the average distance for the new cluster instead of minima or maxima of the alternative approaches\footnote{In the Results section, we also provide two alternative clustering linkage methods -- complete \cite{Sorensen1948} and Ward's \cite{Murtagh2014} algorithms.}.

The resulting hierarchical structure is standardly reported in the form of a dendrogram -- a tree diagram (from Greek \textit{dendron} for a tree and \textit{gramma} for drawing). The diagram shows the most important connections as an output of the clustering analysis and a possible hierarchical structure in the analyzed dataset. The most standard way of drawing a dendrogram is in a form of roots of a tree which is mostly of a rectangular shape. For large datasets and mainly a large number of time series, alternative methods can be utilized for a better graphical representation\footnote{The analysis has been performed in RStudio 0.98.1028 via R 3.1.1 using the \textit{ape} and \textit{stats} packages.}.

\section{Data}

We examine the corruption clustering using the Freedom from Corruption (COR) index of the Heritage Foundation between years 1996 and 2014. Only the countries with statistics for each year within the analyzed period are included in the final dataset. This accounts for 134 countries. The Freedom from Corruption is based primarily on the Corruption Perception Index (CPI)  constructed by the Transparency International. The CPI is a 0-10 range index with 0 meaning the most corrupt country and 10 the opposite. The COR is constructed by simply multiplying the CPI by 10. Hence the COR is on 0-100 scale. If the CPI index is unavailable for a country, the Heritage Foundation estimates the index using expert information from other reliable sources \footnote{U.S. Department of Commerce, Economist Intelligence Unit, Office of the U.S. Trade Representative and official government publications of each country. Additional information can be found at http://www.heritage.org/index/freedom-from-corruption.}. However, it needs to be noted that the number of countries not covered by the CPI is very low. %In the year 2011 the data for COR had to be estimated using the alternative sources only in case of Belize, Fiji, and Micronesia. 

The CPI is not an index of an objective actual state of corruption but of the perception of corruption in the public sector of a country. The Transparency International uses several sources to derive the index\footnote{E.g. African Development Bank Governance Ratings, Economist Intelligence Unit Country Risk Ratings, IMD World Competitiveness Yearbook, Political Risk Services International Country Risk Guide or World Justice Project Rule of Law Index.}. The institutions ask country experts to assess the corruption level in the country in many aspects. The data is then gathered and the CPI is constructed to reflect the level of corruption in a country perceived by its experts.

Of course, the CPI has some natural weaknesses. The most important one is that the CPI is derived from the experts' judgement. The results can be obviously biased by the ``elite'' position of the respondent. The perception of corruption could be different if the ``ordinary ones'' were asked. Another problem lays in the comparison of the CPI scores between years. Temporal changes in ranking can be caused by changes in the country CPI sample or methodology. To overcome these shortcomings, several alternative corruption indices have been developed (e.g. the Global Corruption Barometer, the Bribe Payers Index or the WB's Worldwide Governance Indicators -- Control of Corruption index). Even though all the criticism points to the relevant CPI shortcomings, the index is still the key measure of corruption level across countries. The discussion about CPI validity shows difficulties concerning the corruption measurement which can never be objective and must somehow rely on perceptions. The CPI still offers the largest database of corruption indicator with sufficient historical time series and it remains the most suitable resource for cross-country corruption studies. 

\section{Results}

We apply the average linkage clustering technique based on the Euclidean distances to find a possible hierarchical structure between corruption levels of 134 studied countries. In Fig. \ref{fig1}, we present the final depiction of clustering between analyzed states\footnote{In the Appendix, we also present Figs. \ref{fig2} and \ref{fig3} which are based on the complete and Ward's linkage algorithms, respectively. Even though the details of the hierarchical structure are not identical, the interpretation of the results is qualitatively very close to what we present for the average linkage clustering in the main text. The results are thus quite robust to the linkage algorithm selection.}. Due to a large number of the examined countries, we opt for a spherical representation of the dendrogram. In the centre of the figure, the distance between clusters is the highest and the further we get from the centre of the circle, the closer the clusters or countries are. 

We observe that the network of countries breaks down into four well-defined clusters\footnote{The definition of clusters is quite arbitrary here and we label the main clusters based on a visual inspection. If desirable, the number of clusters could be reduced to two -- Clusters \#1 and \#2 together, and Clusters \#3 and \#4 together. The main message of the results would remain very similar. Either way, the reported results, their structure and implications presented in this section show that such cluster selection is reasonable.}. To help reader with orientation in Fig. \ref{fig1}, we separate the clusters by additional dashed lines. Furthermore, we list the countries according to their cluster membership in Table \ref{tab1} as well. A very interesting clustering structure emerges.

Cluster \#1 is mainly formed of developed countries, i.e. the USA, Japan, the Western states of the EU, Hong Kong, Singapore, Australia and New Zealand, and it forms a solid piece of network structure with the average COR of $82.92\pm1.82$. Members of this cluster thus have very low corruption levels. Compared to the values of other clusters, which are reported in Table \ref{tab1}, the first cluster is very well separated from the others and a considerable distance to other clusters is evident. 

The cluster is clearly separated from the others not only in terms of the corruption level but also in terms of the economic development. The average GDP per capita\footnote{Gross domestic product (GDP) for year 2012 is reported. GDP is a widely used measure of overall economic performance of a given country.} in the group is 52,138 current USD\footnote{Data source: World Development Indicators, WB, year 2012.}. There is no country from the African continent. Both most developed countries from the North and Central America (the USA and Canada) are members of the cluster as well as is the only one and also the most developed country from the Latin America (Chile), the most developed states from Oceania (New Zealand and Australia), except for Kuwait the most developed nations on the Asian continent (Singapore, Japan, Hong Kong and UAE) and 13 most developed countries of Europe. The lowest GDP per capita is less than 7 times lower than the highest one. In other words, the cluster of the least corrupt countries is composed of the most developed nations across all continents. Hence we see that low corruption is clearly connected with economic development. 

Cluster \#2 is more heterogeneous as it contains various EU states such as Malta, Cyprus, Spain, Portugal and Slovenia (i.e. somewhere between the most developed Western states and the post-communist East) as well as a mix of Taiwan, Israel, Barbados, Botswana and others for which the connection can be seen usually only for small groups (such as a mini-cluster of oil exporters Oman, Bahrain and Kuwait). The average COR of $59.21\pm 1.47$ still suggests a decent level of corruption.

This cluster has a lower average GDP per capita of about half the GDP of the most developed cluster (23,521 USD). It consists of the fourth most developed African state (Botswana), the third and fifth most developed countries of the Central and North America, and the second most developed country of the Latin America. Also five (including the first one) from the eleven most developed states in Asia are members of the cluster. The ratio between the lowest and the highest GDP per capita is less than 8. The cluster hence contains states which can be labeled as the upper-middle class of economic development. They exhibit significantly lower economic performance in terms of per capita GDP but still much higher than the other states. These countries are the ``runner-up'' leaders on their continents and they are trying to catch up the leading group. Hence we can regard the cluster as a ``successfully transitioning'' or a ``catching up'' group. The cluster is not only the second richest one but it also has the second lowest corruption level.

Cluster \#3 is then the biggest one with 60 members and the average COR of $24.15\pm 0.78$, i.e. the lowest value among clusters suggesting the highest level of corruption. The basket of countries is quite diverse here ranging from the new EU members (Bulgaria and Romania) to Russia and countries in between (Belarus, Moldova, the Ukraine) to China and India as well many African, Asian and a few Latin American countries. Several interesting implications arise from this structure. First, Bulgaria and Romania seem to be institutionally very different from the rest of the EU as no other EU country falls into the most corrupt cluster. Even the newest EU member state (Croatia) is a member of the last cluster which possesses lower levels of corruption. Second, all BRIC countries (Brazil, Russia, India and China), which historically form a group of large, fast-growing economies, belong to the cluster. These countries are thus still well-behind the developed countries with respect to corruption environment and behavior. And third, various authoritarian regimes fall into this cluster as well -- Azerbaijan, Iran, Laos, North Korea, Syria, Turkey, Venezuela and Vietnam. The corruption levels are thus tightly connected to the form of reign in a country. Note that most of the countries in the least corrupt cluster are standard, western-type democracies. 

From the GDP per capita point of view, the third cluster is again quite diverse. The average GDP per capita is the lowest one with 3,888 USD. However, a range between the richest and the poorest country in the cluster is immense with a ratio of 53 -- between Malawi (287 USD) and Russia (14,090 USD). %The heterogeneity in the economic performance can be seen also at the ``economic rankings'' on the continents. Few countries from the cluster are in the upper half or in the middle. However, the bottom of each continent is composed of countries which are members of this clusters. E.g. on African continent the members of the cluster are the most developed country (Libya) \footnote{We use GDP per capita data from year 2012. It is very likely that Libya would have different ranging at the moment because of the dismal security situation} but also ten the least developed states. The cluster contains also the least developed countries from Central and North American, South American, Asian or European continents. 
Nonetheless, we can regard the cluster as a group of countries with the lowest economic performance. The cluster contains not only the poor countries or countries in civil war situations (e.g. Libya or Syria) but also transition states with a high future economic potential (e.g. Brazil or Turkey). It is likely the highest corruption can be a relevant obstacle for the development of these countries and hence the corruption-economic development correlation seems to be valid even in this case.

Cluster \#4 is somewhere between the previous two clusters with the average COR of $41.28\pm 0.98$. This is again quite a heterogeneous cluster (the ratio between the the richest and the poorest country is 45) with several interesting subgroups. First, the whole Visegrad group (the Czech Republic, Hungary, Poland and Slovakia), post-communist countries in the Central Europe, as well as the Baltic post-communist states (Latvia and Lithuania) are members of this cluster\footnote{Estonia is not in the dataset due to incomplete time series.}. Second, the newest member state of the EU -- Croatia -- is also part of the cluster and it thus readily joins the post-communist countries of the region, overtaking Bulgaria and Romania. Third, some of developed countries such as Italy and Greece fall into this cluster as well which might give a notion about economic problems and its causes in these states. And fourth, practically the whole region of the North Africa (Algeria, Egypt, Morocco and Tunisia) is inside the cluster. From the economic perspective, the last cluster falls within the second and the third one with the average GDP of 9,751 USD.  The connection between corruption and economic development is thus supported even by the last detected cluster.

%The fourth cluster is in the middle between the second and third cluster also in terms of the average GDP per capita (9,751 USD). The cluster contains the economic African leaders, "lower middle" countries following the first and second cluster states on the Central and North American, Asian and European continent. There is only one country from the South American continent (Argentina) which is the third most developed one.

%The fourth cluster exhibits not catastrophic corruption or economic level however we see that the GDP per capita is more than twice lower compared to the second cluster of successful transition countries. Hence the countries in the fourth cluster are more at the beginning of their journey to stable economic performance than in the middle.

\section{Conclusions}

We have inspected a possible clustering structure of the corruption perception among 134 countries. Using the average linkage clustering, we have uncovered a well-defined hierarchy in the relationships among countries. Four main clusters have been identified and they suggest that countries worldwide can be quite well separated according to their perception of corruption. Moreover, we have found a strong connection between corruption levels and a stage of development inside the clusters. The ranking of countries according to their corruption perfectly copies the ranking according to the economic performance measured by the gross domestic product per capita of the member states. Even though our analysis does not (as it cannot) discuss potential causality between the two, the reported results remain promising. We believe that our analysis, which is the first one to the best of our knowledge, can be used as a starting point of various analyses in social sciences which struggle to find an appropriate mechanism of separating countries according to such a specific criterion as corruption. We are convinced that the appropriate clustering can be beneficial especially for studies concerning international trade where the corruption is gaining increasing attention and the studies lack rigorous differentiation of corrupt countries.

\section*{Acknowledgements}

This research has been supported by the CERGE-EI Foundation under a program of the Global Development Network with a grant number RRC15+55. All opinions expressed are those of the authors and have not been endorsed by CERGE-EI or the GDN. The research has also received funding from the Czech Science Foundation under project No. P402/12/0982 and the SVV project 260 113 ``Strengthening Doctoral Research in Economics and Finance''. Michal Paulus acknowledges the support of the Deutscher Akademischer Austausch Dienst (DAAD) sponsoring the long-term research visit (grant No. 50015537) at the Humboldt Unviersität zu Berlin and of the Grant Agency of the Charles University in Prague. 

%\newpage

\section*{References}
\bibliographystyle{unsrt}
\bibliography{Corruption}

\begin{figure}[htbp]
\center
\begin{tabular}{c}
\includegraphics[width=6in]{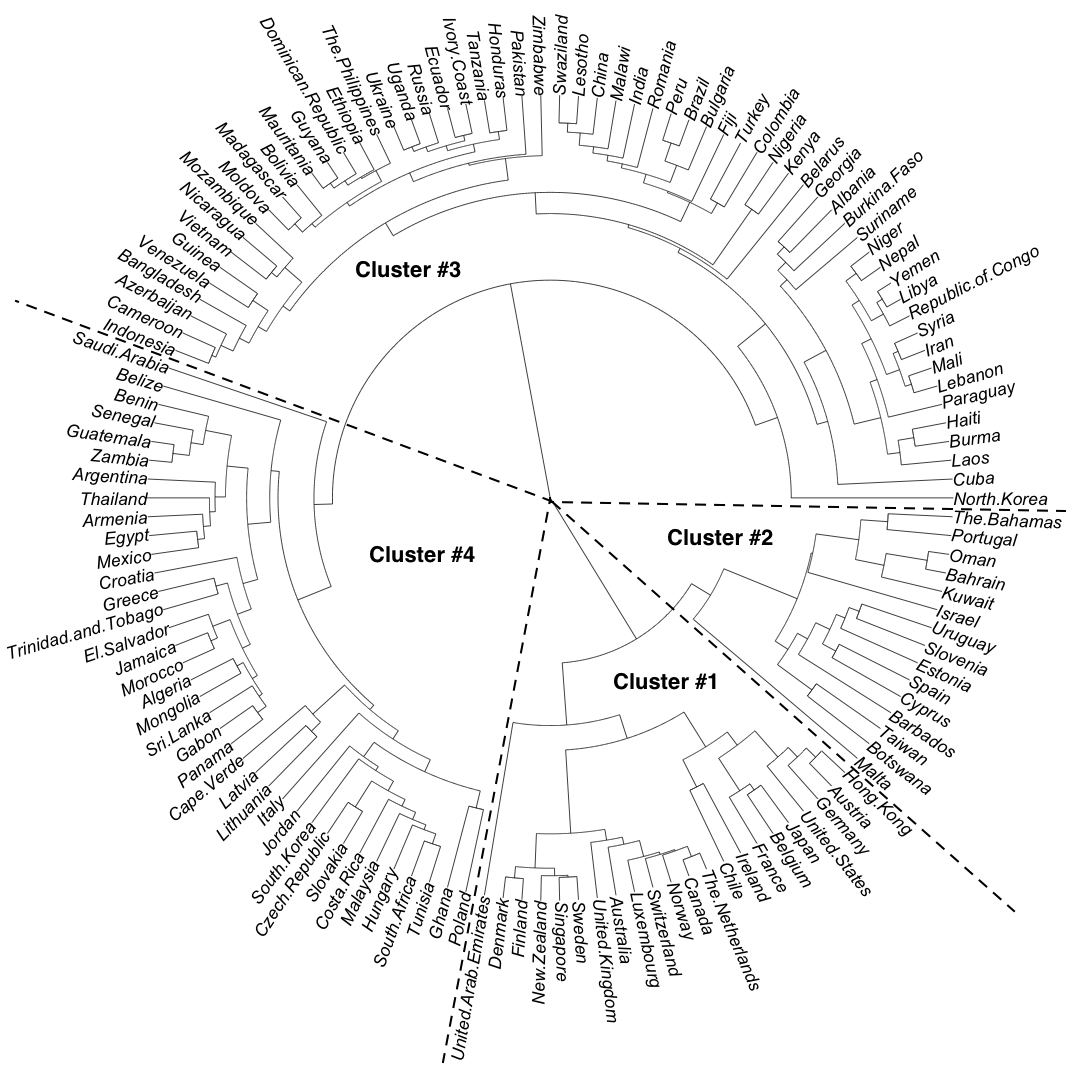}\\
\end{tabular}
\caption{\footnotesize\textit{Dendrogram of corruption perception.} The dendrogram is based on the average linkage clustering procedure using the Euclidean distance as a measure of dissimilarity. The spherical depiction gives a base of the tree in the centre and roots are growing centrifugally. The identified clusters are based on a visual inspection and for a better orientation, these are separated by dashed lines.  \label{fig1}}
\end{figure}

\begin{table}[htbp]
\centering
\caption{\textbf{Countries in clusters.} Clusters from Fig. \ref{fig1} are listed here and the average values with standard errors are reported as well. The countries are listed alphabetically within the clusters.}
\label{tab1}
\footnotesize
\begin{tabular}{ccccc}
\toprule \toprule
\multicolumn{5}{c}{Cluster \#1 -- average $82.92 \pm 1.82$}\\
\midrule
Australia&Austria&Belgium&Canada&Chile\\
Denmark&Finland&France&Germany&Hong Kong\\
Ireland&Japan&Luxembourg&Netherlands&New Zealand\\
Norway&Singapore&Sweden&Switzerland&United Arab Emirates\\
United Kingdom&USA&&&\\
\midrule
\multicolumn{5}{c}{Cluster \#2 -- average $59.21 \pm 1.47$}\\
\midrule
Bahamas&Bahrain&Barbados&Botswana&Cyprus\\
Estonia&Israel&Kuwait&Malta&Oman\\
Portugal&Slovenia&Spain&Taiwan&Uruguay\\
\midrule
\multicolumn{5}{c}{Cluster \#3 -- average $24.15 \pm 0.78$}\\
\midrule
Albania&Azerbaijan&Bangladesh&Belarus&Bolivia\\
Brazil&Bulgaria&Burkina Faso&Burma&Cameroon\\
China&Colombia&Cuba&Dominican Republic&Ecuador\\
Ethiopia&Fiji&Georgia&Guinea&Guyana\\
Haiti&Honduras&India&Indonesia&Iran\\
Ivory Coast&Kenya&Laos&Lebanon&Lesotho\\
Libya&Madagascar&Malawi&Mali&Mauritania\\
Moldova&Mozambique&Nepal&Nicaragua&Niger\\
Nigeria&North Korea&Pakistan&Paraguay&Peru\\
Philippines&Republic of Kongo&Romania&Russia&Suriname\\
Swaziland&Syria&Tanzania&Turkey&Uganda\\
Ukraine&Venezuela&Vietnam&Yemen&Zimbabwe\\
\midrule
\multicolumn{5}{c}{Cluster \#4 -- average $41.28 \pm 0.98$}\\
\midrule
Algeria&Argentina&Armenia&Belize&Benin\\
Cape Verde&Costa Rica&Croatia&Czech Republic&Egypt\\
El Salvador&Gabon&Ghana&Greece&Guatemala\\
Hungary&Italy&Jamaica&Jordan&Latvia\\
Lithuania&Malaysia&Mexico&Mongolia&Morocco\\
Panama&Poland&Saudi Arabia&Senegal&Slovakia\\
South Africa&South Korea&Sri Lanka&Thailand&Trinidad and Tobago\\
Tunisia&Zambia&&&\\
\bottomrule \bottomrule
\end{tabular}
\end{table}

\newpage

\section*{Appendix}

\begin{figure}[htbp]
\center
\begin{tabular}{c}
\includegraphics[width=6in]{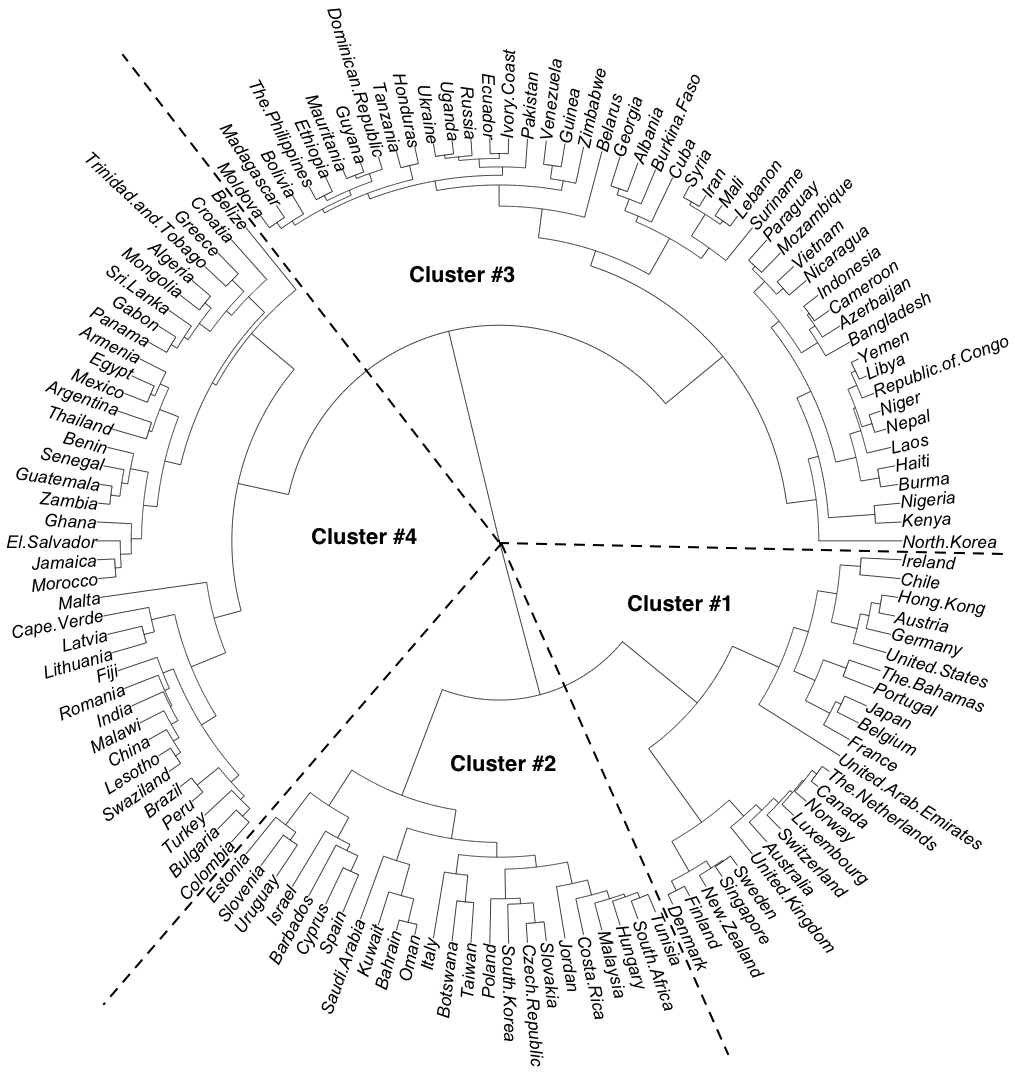}\\
\end{tabular}
\caption{\footnotesize\textit{Alternative dendrogram of corruption perception -- complete linkage.} The dendrogram is based on the complete linkage clustering procedure using the Euclidean distance as a measure of dissimilarity. The spherical depiction gives a base of the tree in the centre and roots are growing centrifugally. The identified clusters are based on a visual inspection and for a better orientation, these are separated by dashed lines. Structure of the clusters is very similar to the one based on the average linkage clustering in Fig. \ref{fig1}. \label{fig2}}
\end{figure}

\begin{figure}[htbp]
\center
\begin{tabular}{c}
\includegraphics[width=6in]{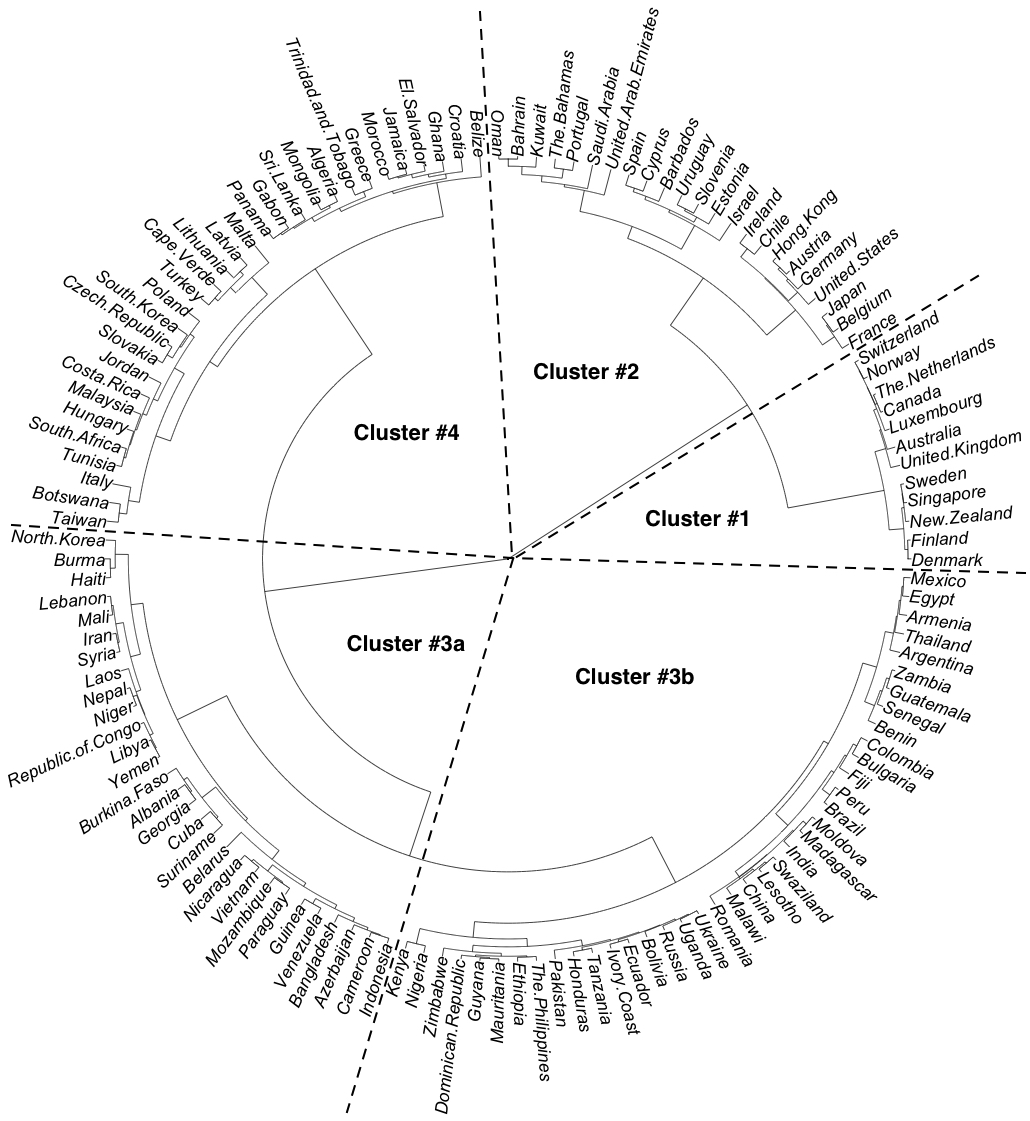}\\
\end{tabular}
\caption{\footnotesize\textit{Alternative dendrogram of corruption perception -- Ward's linkage.} The dendrogram is based on the Ward's \cite{Murtagh2014} linkage clustering procedure using the Euclidean distance as a measure of dissimilarity. The spherical depiction gives a base of the tree in the centre and roots are growing centrifugally. The identified clusters are based on a visual inspection and for a better orientation, these are separated by dashed lines. Structure of the clusters is more complex than for the average and complete linkage methods, such as splitting of Cluster \#3 into two clusters. Nevertheless, the most important features of the network remain intact. \label{fig3}}
\end{figure}

\end{document}